# Moiré Potential, Lattice Relaxation and Layer Polarization in Marginally Twisted MoS$_2$ Bilayers.

*Nikhil Tilak[1], Guohong Li[1], Takashi Taniguchi[2], Kenji Watanabe[2], Eva Y. Andrei[1*]*

1 Department of Physics and Astronomy, Rutgers, The State University of New Jersey

2 National Institute for Materials Science, Tsukuba, Japan.

ABSTRACT:

Artificially twisted heterostructures of semiconducting transition metal dichalcogenides (TMDs) offer unprecedented control over their electronic and optical properties via the spatial modulation of interlayer interactions and structural reconstruction. Here we study twisted MoS$_2$ bilayers in a wide range of twist angles near 0° using Scanning Tunneling Microscopy/ Spectroscopy. We investigate the twist angle-dependence of the moiré pattern which is dominated by lattice reconstruction for small angles (<2°) leading to large triangular domains with rhombohedral stacking. Local spectroscopy measurements reveal a large moiré-potential strength of 100-200 meV for angles <3°. In reconstructed regions we see a bias-dependent asymmetry between neighboring triangular domains which we relate to the vertical polarization which is intrinsic to rhombohedral stacked TMDs. This viewpoint is further supported by spectroscopy maps and ambient Piezoresponse measurements. Our results provide a microscopic perspective on this new class of interfacial ferroelectrics and can offer clues for designing novel heterostructures which harness this effect.

**Introduction:**

The properties of layered Van der Waals (vdW) materials stacked on top of each other to create vertical heterostructures which can be remarkably different from the constituent materials. The interlayer twist angle, can be precisely controlled in these heterostructures[1] to create so-called moiré materials[2]. One of the most well-studied, magic angle twisted Bilayer Graphene (MA-TBG)[3], exhibits a variety of correlated states including superconductivity, Chern insulators, and orbital Ferromagnets[4-11].

Recently, twisted bilayers of semiconducting Group VI Transition Metal Dichalcogenides (TMDs) of the form $MX_2$ where M∈ (Mo, W) and X∈ (S, Se, Te) have gained tremendous interest. Unlike Graphene, twisted TMDs host flat bands near the Valence band edge for a wide range of twist angles. The band width of these flat bands can, in theory, be made arbitrarily small by reducing the twist angle. Emergent Quantum phenomena which arise in twisted TMDs by filling these flat bands were probed using electrical transport[12] and various optical probes[13-17].

2H TMD monolayers aren't true atomic monolayers but consist of a layer of metal atoms sandwiched between two chalcogen layers with a trigonal prismatic unit cell. While bulk 2H TMDs are inversion symmetric, individual monolayers lack inversion symmetry because atoms of different elements occupy different lattice sites. Consequently, an interlayer twist angle $\theta$ is not equivalent to $60° - \theta$. Twisted TMD heterostructures can therefore be categorized into two types: Parallel stacked ($\theta \sim 0°$) or Anti-Parallel stacked ($\theta \sim 60°$). Here $\theta = 60°$ corresponds to 2H stacking while $\theta = 0°$ combined with a lateral shift of $1/3^{rd}$ of a lattice constant corresponds to 3R stacking. The existence of multiple TMD materials allows for the creation of heterobilayers, where two monolayers of different TMDs are stacked together or homobilayers, where two monolayers of the same TMD form the stack.

Several STM experiments[18-24] have studied the topography and local electronic band structure in these systems. Most of these studies have focused on heterobilayers and/or relatively large twist angles. However, an STM characterization of twisted TMD homobilayers with very small twist angles is lacking. Such marginally twisted homobilayers are expected to undergo significant lattice reconstruction and form large areas of perfect R stacking. While the 3R bulk polytype of $MoS_2$ exists in nature, it is exceedingly rare. Marginally twisted TMDs combined with a local probe like STM provides a simple platform for studying this rare polytype. Furthermore, due to broken centrosymmetry, R stacked regions are also expected to host interfacial ferroelectricity [25, 26].

In this work we focus on Parallel-stacked $MoS_2$ homobilayers with a range of twist angles between 4.9° and 0.28°. $MoS_2$ was chosen since it's perhaps the most well studied TMD and is relatively easy to exfoliate. We expect the main observations to be generally true for all other TMD homobilayers with minor differences stemming mainly from differences in the strength of spin-orbit coupling.

We studied 3 devices with different twist angles in a homebuilt STM system[27, 28] at 77 K (Figure 1a) (see methods for details). Two distinct device geometries were employed in this study (Figure 1a right). For samples A and B with relatively large, target twist angles of 3° and 2° respectively, a monolayer graphene (MLG) was first transferred on the bottom hBN flake and the twisted $MoS_2$ stack was built on top of it for better electrical contact at low temperatures. For the sample C with a target twist angle of 1°, the twisted MoS2 stack was fabricated on bare hBN, and was partially contacted by a small flake of MLG to reduce the contact resistance. This geometry was chosen so that a large, out-of-plane displacement field could be applied using the Si backgate. While the contact worked at room temperature, it showed a rectifying behavior at 77 K which made scanning at a positive sample bias (conduction bands) unreliable in the regions not covered with graphene. Nevertheless, we were able to reliably scan the sample at negative biases providing access to states in the valence bands. In all three devices, an electrode with a capacitance-sensing geometry was used to navigate the STM tip to the micron sized heterostructure without optical access[29].

**Results and Discussion**

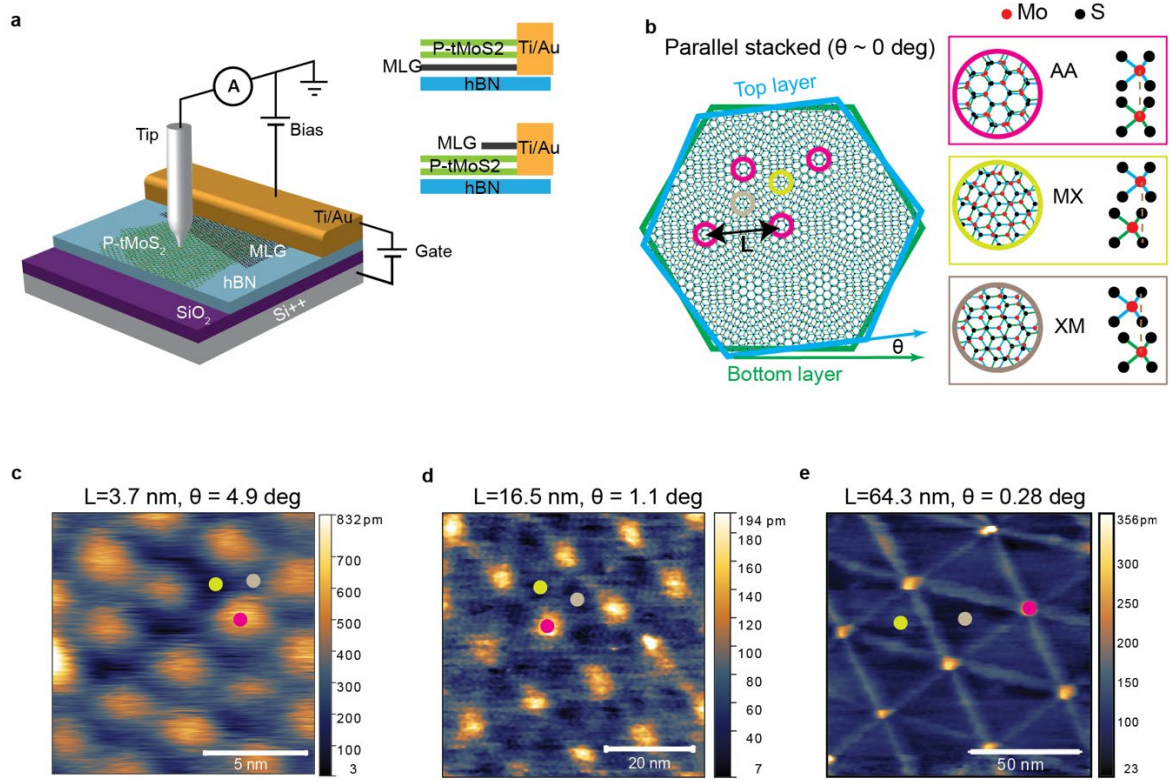

**Figure 1.** Measurement setup and structure of Parallel stacked twisted MoS$_2$ **a.** STM measurement setup and device structure. Two distinct kinds of device geometries were employed. One where a monolayer Graphene was placed between the twisted MoS$_2$ stack and hBN and another one where a monolayer graphene flake partially contacted the twisted MoS$_2$ stack from the top. **b** The moiré lattice structure of Parallel-stacked (twist angle near 0°) twisted MoS$_2$. Zoomed views of selected high-symmetry locations in the moiré unit cell labelled AA, MX and XM are shown along with a side-view to show atomic registrations in the top and bottom MoS$_2$ monolayers. Three representative STM topography scans with decreasing twist angle are shown in panels **c-e**. The high symmetry point locations in each scan are indicated by the colored dots. The tunneling parameters are (Bias = -2 V, Current=40 pA) for **c**, (Bias = -2.5 V, Current=70 pA) for **d** and (Bias = -2.1 V, Current=20 pA) for **e**. Regions **c** and **d** are exposed P-tMoS2 while region **e** is covered with MLG. All scans were recorded at 77 K.

Figure 1b shows a schematic of a moiré superlattice formed by P-stacking two rigid MoS$_2$ monolayers with an interlayer twist. For homobilayers, the moiré wavelength L depends on the twist angle $\theta$ and the lattice constant of the monolayers $a$ as $L = a/(2 \sin(\theta/2))$. Although the relative atomic registration between the two monolayers varies continuously in space, three high-symmetry stacking-locations can be identified within every moiré unit-cell (Figure 1b right). In the AA sites (magenta circle) both the Mo and S atoms in the top layer are aligned with the corresponding atoms in the bottom layer. The sites labelled MX (green circle) and XM (brown circle) are two versions of Rhombohedral stacking arrangement. In the MX (XM) sites the Mo(S) atoms in the top layer are aligned with the S(Mo) atoms in the bottom layer while the S(Mo) atoms in the top layer sit on empty sites in the bottom layer.

Three such Moiré superlattices with increasing moiré wavelengths are shown in STM topography scans (Figure 1c-e). For P-stacked TMDs, Density Functional Theory (DFT) calculations[30, 31] have shown that the inter-layer separation (ILS) at the AA sites is larger than the MX and XM sites by about 60 pm due to strong repulsion of the out-of-plane S-$p_z$ orbitals. The MX and XM sites have the same ILS. The stacking energy closely follows the ILS. Following this, we identify the bright circular spots which appear most intense in topography with the AA sites and the 6 surrounding dips as alternating MX and XM sites. The average moiré wavelength was measured between the center of neighboring AA spots and the local twist angles were inferred assuming $a = 0.316$ nm for 2H-MoS$_2$.

For relatively large twist angles, this simple model where only the ILS is different for different stackings, but the atomic lattice is perfectly rigid in-plane can explain the observed topography scans. If we extrapolate this rigid model to smaller twist angles, the size of the AA, MX and XM sites should continuously increase in proportion to the expanding moiré cell area with decreasing twist angles. But as noted above, the MX and XM sites, have a lower stacking energy per unit area than the AA sites and it was predicted[32] that as the twist angle decreases it becomes energetically favorable to increase the relative area of the MX/XM sites and shrink the AA sites at the cost of in-plane deformation of the atomic lattices. This lattice strain results in narrow domain walls [30] which terminate at the AA sites and separate large triangular MX/XM sites which have almost perfect R stacking. This results in a triangular tiling pattern which was reported in a recent SEM study[33]. Analogous lattice reconstruction effects have been previously observed in marginally twisted Bilayer Graphene (SA-TBG)[34].

Evidence of lattice reconstruction can be seen in the topography scans in Figure 1d-e. We found that the size of the AA sites saturates to about 6 nm for twist angles below ~2°. Domain walls can be seen in Figure 1d where the twist angle is 1.1° and even more clearly in Figure 1e where the twist angle is 0.28°. Additional external strain arising from wrinkles and bubbles which form during sample fabrication, can further distort the moiré lattice due to its magnifying effect[35]. In sample C, we have recorded some such large, distorted moiré patterns (Supplementary Figure 2).

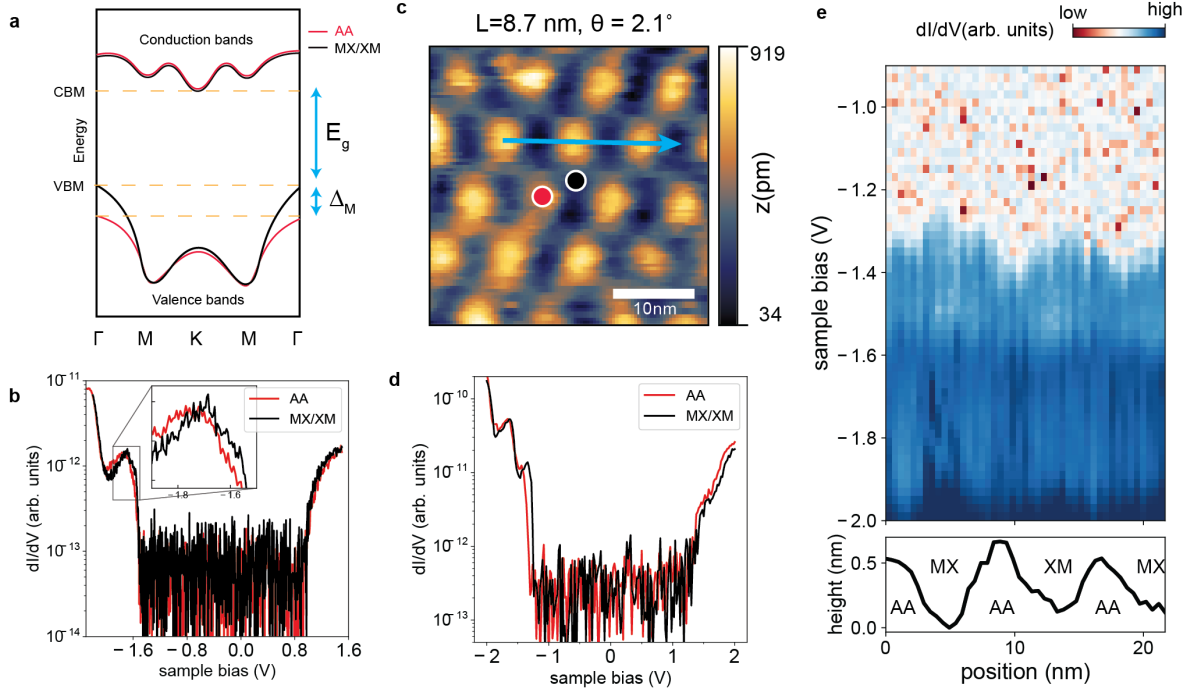

**Figure 2.** Moiré potential induced band oscillations in twisted MoS$_2$ **a.** A schematic of the band structure of tMoS2 for isolated AA (red) and MX/XM (black) sites. The CBM is located around the K (or Q) point and is approximately the same for AA and MX/XM sites. The VBM is located at the $\Gamma$ point. The VBM is expected to be lower at the AA stacking sites than the MX/XM sites by an amount $\Delta_M$ which is a measure of the moiré potential strength. **b.** The dI/dV spectra measured at the AA and MX/XM sites of a 4.9° twist angle region. The inset shows a zoom-in around the VBM which indicates a small downward shift of the VBM at the AA sites and an additional density of states at the MX/XM regions near the VBM compared to the AA region. **c.**

STM topography scan of a 2.1° twisted MoS2 region measured at Bias = -1.7 V and Current=200 pA. **d.** dI/dV spectra measured at an AA site and a MX/XM site with positions indicated in c with colored dots. The Valence band edge in the AA region is shifted to lower energy by about 100 meV compared to the MX/XM regions. The Conduction band edge is also shifted lower for AA sites compared to MX/XM sites but by a much smaller amount. **e.** The top panel shows the evolution of the dI/dV spectra as a function of bias and position along the blue arrow in **c.** The bottom panel shows the simultaneously acquired STM topography. The VBM oscillates inversely with the topography, reaching the lowest energy at the AA sites and the highest energy at the MX/XM sites by about 100 meV.

Band structure calculations[30, 31, 36] have shown that for P-tMoS2 the moiré superlattice acts as an additional long-wavelength periodic potential which leads to the formation of several moiré minibands near the Valence band edge. A schematic of the expected band structures for isolated AA and MX/XM stackings is shown in Figure 2a. The valence band maximum (VBM) is located at the Γ point and is lower at the AA sites than the MX/XM sites by an amount $\Delta_M$. This is a direct consequence of the higher ILS in the AA sites leading to a lowering of the hybridization strength. As a result, the lowest energy holes are localized in a hexagonal lattice formed by the MX/XM sites. On the other hand, the conduction band minimum (CBM) is less sensitive to stacking.

First, we focus on the dI/dV spectroscopy measured on areas with local twist angle greater than 2°. The spectra measured on the area with twist angle of 4.9° (Figure 2b) show an average band gap of ~ 2.1 eV which is close to the gap measured in MoS2 monolayers. The conduction band edge is located around +0.8 V for both the AA and MX/XM sites. There is a broad peak centered around -1.7 V at both the AA and MX/XM sites. On closer inspection, there is an additional density of states in the MX/XM region near the Valence band edge compared to the AA sites (Figure 2b inset). We identify this with the contribution of a moiré miniband maximum centered around the Γ-point in the mini-Brillouin Zone (Figure 2a).

On decreasing the twist angle to about 2.1° (Figure 2c) noticeable differences are seen in the Valence band edge in the spectra measured on the AA and MX/XM sites (Figure 2). In all the spectra, two broad peaks can be seen near the Valence band maximum around -1.5 V and -1.7 V. The intensity of the peak at -1.5 V is higher at the MX/ XM sites. The Valence band maximum (VBM) at the AA sites was found to be shifted to lower energies compared to the MX/XM sites. The conduction band edge is also shifted lower but by a much smaller amount. This observation matches well with the predicted[30, 31, 36] band structure of P-tMoS$_2$.

To further investigate this energy shift we measured dI/dV spectra at points along the cyan arrow shown in Figure 2c. Since the shifts are the most pronounced near the Valence band edge, we plot the spectra as a colormap focused near the valence band (Figure 2d top panel). The corresponding topographic height is plotted on the bottom panel. A clear oscillation of the Valence band edge with the topographic height can be seen. The VBM is lowest at the AA sites and highest at the MX/XM sites. The measured oscillation amplitude of ~100 meV gives us an estimate of the moiré potential ($\Delta_M$) for this twist angle. We found that the moiré potential is dependent on twist angle and reaches values upto 200 meV for smaller twist angles. This is lower than the theoretically predicted $\Delta_M$~300 meV[30].The spectra at the MX and XM sites were found to be virtually identical for large twist angles. Differences were however observed for smaller twist angles and are discussed below.

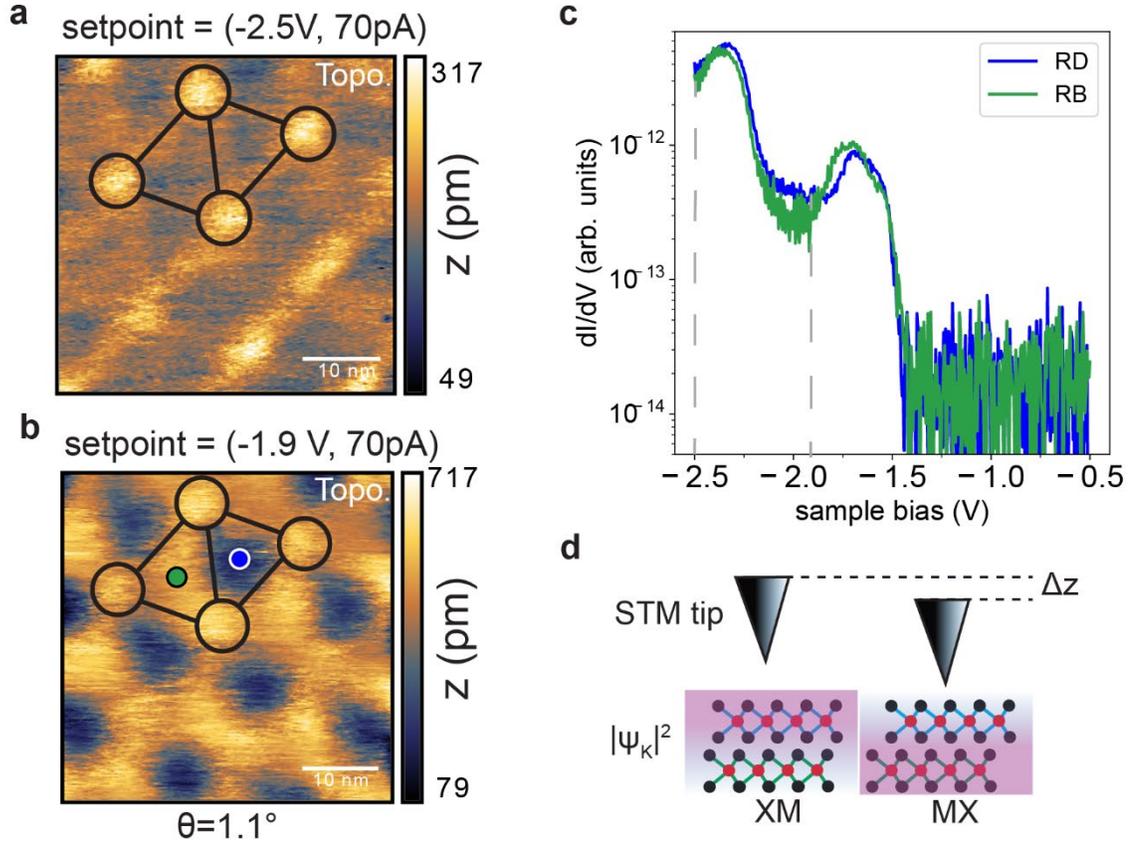

**Figure 3.** Broken mirror symmetry in small twist angle twisted MoS$_2$ STM topography scans of a 1.1° twisted MoS$_2$ region taken at a tunneling current of 70 pA and at a bias setpoint of -2.5 V (panel **a**) and -1.9 V (panel **b**) respectively. There is a small height difference in the MX and XM sites at -2.5 V which dramatically increases at -1.9 V. **c**. Point spectra measured at the center of the RD (dark) and RB (bright) sites with a stabilization set point of (-2.2 V, 20 pA). The bias setpoints for the topography in panel **a** and **b** are indicated by the gray dashed lines. Although the VBM in both regions appears to be the same, there are several differences in the spectra at lower energies. Panel **d** is a cartoon showing the difference in the wavefunction weight centered on the K valley on the top and bottom layer and the movement of the STM tip as it scans over these regions.

For regions with small twist angles the topography was observed to be highly bias dependent. For example, Figure 3a and 3b show two topography scans of the same region ($\theta = 1.1°$) measured at a bias voltage of -2.5 V and -1.9 V respectively with the tunneling current held constant at 70 pA. At a bias setpoint of -2.5 V (Figure 3a) the familiar topography was seen with bright AA sites and narrow domain walls connecting them. Almost no topographic difference was seen in the MX and XM sites. However, on reducing the bias to -1.9 V (Figure 3b), one type of rhombohedral stacking site appears brighter than the other, thus breaking the $C_{2y}$ mirror symmetry of the moiré superlattice. The observed intensity difference between the two is ~0.6 nm which is a significant fraction of the thickness of one monolayer (~0.8 nm). From here on we call the brighter of the two the RB sites (Rhombohedral Bright) while the darker ones the RD sites (Rhombohedral Dark). The topographic intensity difference between the RB and RD sites increases systematically as the bias voltage is moved towards the VBM (see Supplementary Figure. 3). This trend was seen in all areas with twist angles < 2°.

From lattice relaxation models we expect the ILS to be equal for the MX and XM regions. Therefore, the observed topographic height difference is puzzling at first. It is instructive to recall that the measured STM topography is a convolution of the real height variation and the LDOS variation. If there is a real height variation between the MX and XM sites it should be independent of the tunneling voltage setpoint. Additionally, contact mode Atomic Force Microscopy on a comparable sample with a similar triangular moiré pattern, did not reveal any topographic height variation in the MX/XM regions to within 10 pm[33]. We can thus safely rule out a substantial height difference between the MX and XM regions.

This points towards the LDOS variation as the origin of the observed asymmetry. The dI/dV spectra measured at the RD and RB lattice sites (Figure 3c) show that while the VBM is nearly identical for the RD and RB regions, there are differences in the LDOS in the -1.7 eV to -2.0 eV energy range. The LDOS is higher for the RB regions in the -1.9 eV to -1.5 eV range, while it is slightly higher for the RD regions in the -2.4 eV to -1.9 eV range. For flat samples the topographic contrast depends on the integrated LDOS (I-LDOS) within the bias window. For set points near -2.5 V the I-LDOS is almost equal for the RB and RD regions resulting in no contrast, whereas for set points between, -1.9 V and the VBM, the I-LDOS is higher for the RB sites compared to the RD sites. This difference is reflected in the topography signal measured by the STM.

Now we turn to the physical origin of the measured differences in LDOS. MX/XM regions are non-centrosymmetric. The low energy electrons and holes can therefore prefer one layer over another leading to a net interlayer charge transfer resulting in an intrinsic vertical polarization as shown in DFT calculations [15, 37, 38]. The direction of this polarization is opposite for MX and XM stacking. Furthermore, as a result of the effective vertical electric field, the bands near the K-point are almost entirely layer-polarized. This layer polarization is absent for the states near the Γ-point due to strong interlayer hybridization.

We believe that the observed mirror symmetry breaking is a direct consequence of this layer polarization. In brief, since the STM tip is physically closer to the top $MoS_2$ layer it is more sensitive to the LDOS of the top layer than the bottom layer. When the tip is over an XM region (Figure 3d left), which has a higher VB wavefunction weight, the tunneling current increases and as a result the tip is retracted from the surface by the feedback loop to maintain the current setpoint. The exact opposite happens when the tip is over an XM region (Figure 3d right). These observations allow us to infer that the RB(RD) sites have XM(MX) stacking. The states near the VBM are centered at the Γ point, which is expected to have no layer polarization. This explains the identical LDOS near the VBM. As a further confirmation we performed LDOS spatial mapping at a 0.77° twisted region. The topography (Figure 4a) was measured at a setpoint of -1.8 V, 50 pA and shows mirror symmetry breaking. Normalized dI/dV spectra (Figure 4b) measured at the three high symmetry sites in the moiré unit cell show that firstly, the Valence band edge in the AA region is shifted to lower energy compared to the MX/XM regions and secondly, the LDOS at the RB regions is significantly higher than the RD regions between -1.8 V and -1.5 V. This is also reflected in the LDOS map extracted at energy -1.62 eV (Figure 4c) which looks remarkably like the observed topography. The LDOS map at an energy of -1.42 eV (Figure 4d) which is very close to the VBM for the MX and XM regions shows no mirror symmetry breaking as expected from theory. Also note that there is very low LDOS at the AA sites and the domain walls because the VBM is shifted lower there as expected.

While many STM experiments[39, 40] have previously studied ferroelectric 2D materials with an in-plane Polarization, to the best of our knowledge, this is the first work to see signatures of vertical Polarization, albeit indirectly. However, our efforts to manipulate the domains with a vertical displacement field applied using the backgate were unsuccessful and are discussed in the supplementary information.

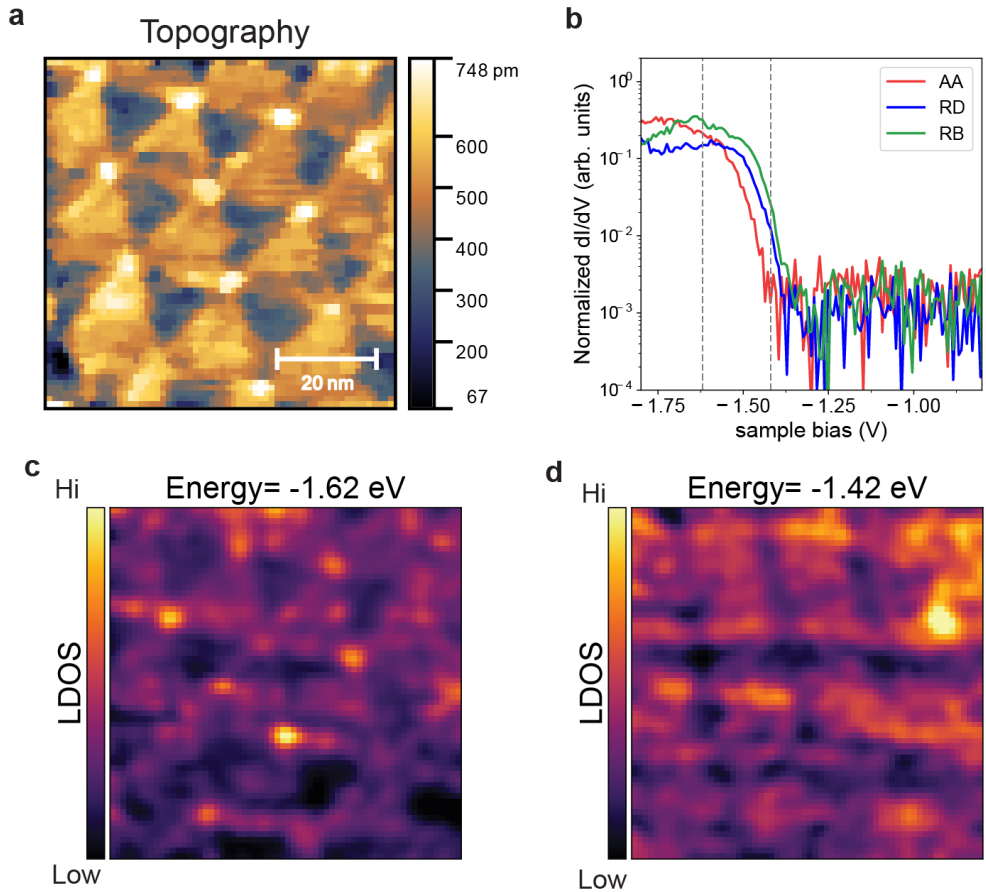

**Figure 4.** LDOS mapping in small twist angle twisted MoS$_2$. **a.** Topography scan of a 0.77° twisted MoS$_2$ region which was acquired at the same time as the LDOS map at a stabilization set point of V$_b$= -1.8 V, I = 50 pA. **b.** Normalized dI/dV spectra measured at the three high symmetry sites in the moiré unit cell. The Valence band edge in the AA region is shifted to lower energy compared to the MX/XM regions. The LDOS is higher at the RB sites compared to the RD sites. Panels **c** and **d** are the LDOS maps extracted at the energies -1.62 eV and -1.42 eV respectively. The map at -1.62 eV shows a high LDOS at the RB regions compared to the RD regions. The map at -1.42 eV shows an almost equal LDOS at both RD and RB sites and a negligible LDOS at the AA and Domain wall sites.

As an independent check of our results, we used RE-PFM [41] to study a P-tMoS$_2$ device with a 0.2° twist angle (Figure 5a). While the topography scan (Figure. 5b) is featureless, the PFM Magnitude (Figure 5c) and Phase (Figure 5d) channels show a clear moiré pattern. The moiré pattern is distorted, presumable due to the wrinkles outside the field of view (see Supplementary Figure 5). As the moiré period increases towards the top-center of the image area, clear triangular domains with alternating contrast can be seen in both the Magnitude and Phase signal. Since there are no significant strain gradients in the center of the triangular domains, this contrast cannot be explained by higher order effects such as flexoelectricity. We therefore attribute this contrast to the Piezoelectric or electrostatic effect in this system. Note that similar results were obtained earlier in a marginally twisted MoSe$_2$ homobilayers [25]. These results also complement previous Conductive AFM measurements of P-tMoS2 which focused on states near the CBM[26].

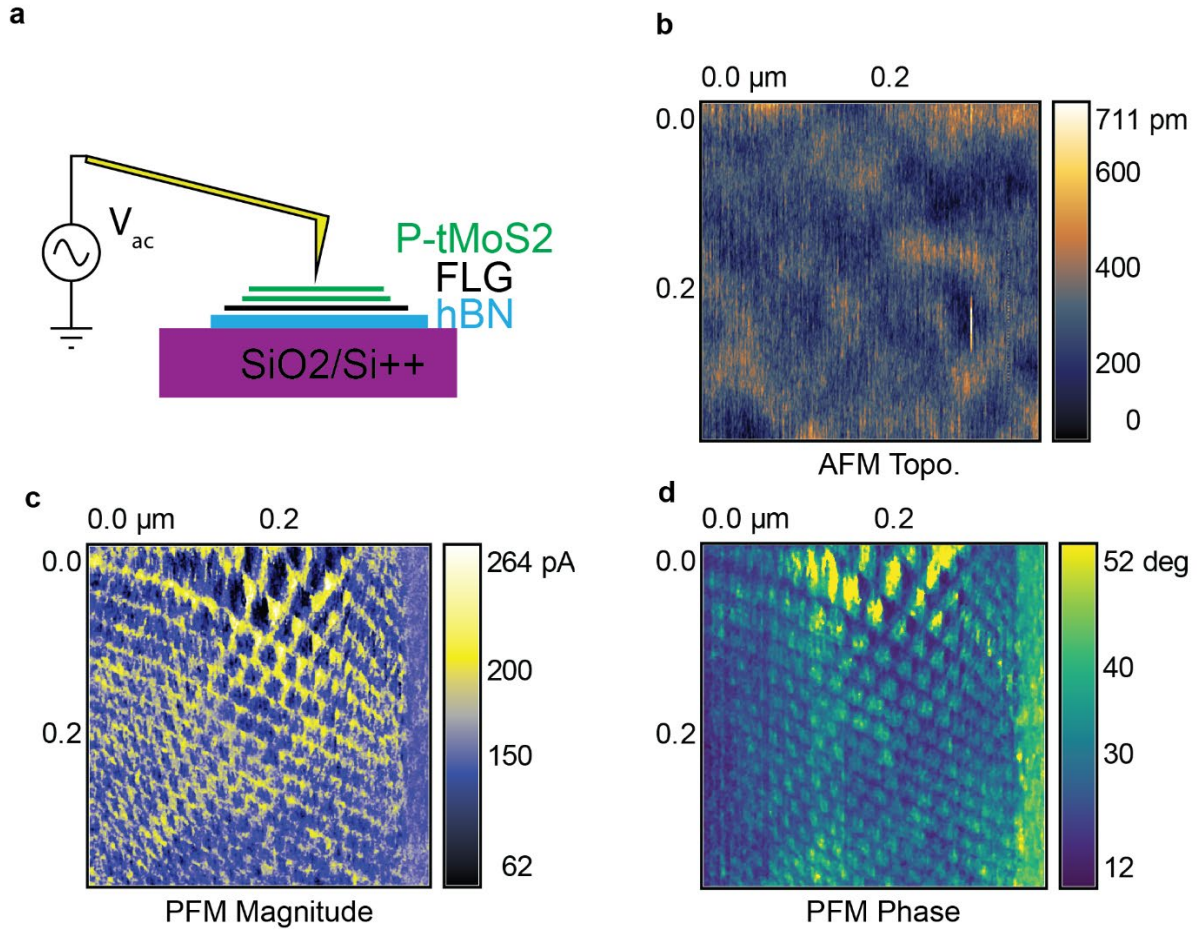

**Figure 5**. RE-PFM measurements for a P-tMoS2 device. **a.** Shows an illustration of the RE-PFM measurement setup. **b.** Contact mode topography scan of a 400 $nm^2$ area. **c** and **d** are the Magnitude and Phase of the vertical deflection signal showing a moiré pattern with varying twist angles. Triangular domains with alternating contrast emerge as the moiré wavelength increases in the top-center region of the image area.

**Summary**

In summary, we have presented a systematic study of the structure and electronic properties of Parallel twisted MoS2 homobilayers using STM/STS. The moiré potential strength was found to be 100-200 meV from STS measurements for twist angles < $3^0$. At marginal twist angles, large triangular domains with perfect R stacking separated by narrow domain walls were observed. Furthermore, a bias dependent breaking of mirror symmetry was observed between the MX/XM regions which we relate to the intrinsic Polarization of R stacked TMDs. Our findings are also complemented by ambient RE-PFM measurements. These results can guide the development of novel heterostructures which can harness this new class of interfacial Ferroelectrics for applications.

**Methods**

Sample preparation:

The devices were made using a tear and stack method. Bulk 2H-MoS$_2$ flakes (SPI supplies) were exfoliated using adhesive tape and transferred onto a pre-made polymer stack consisting of Poly methyl methacrylate (PMMA) and a sacrificial layer of Poly Vinyl Alcohol (PVA). The PMMA film was detached from the substrate (Silicon wafer) and mounted under an optical microscope. Flakes of monolayer MoS$_2$ were identified using optical contrast. A 1mm$^2$ piece of Poly-dimethyl Siloxane (PDMS) was placed behind the MoS$_2$ flake and the whole stack was transferred to a glass slide to make a transfer-handle.

Hexagonal Boron Nitride (hBN) was exfoliated onto a p-doped Silicon chip with a 285 nm SiO2 capping layer and annealed at 350 C in air for 6h to remove tape glue residues. In devices A and B, a large flake of single layer Graphene (MLG) was transferred onto the hBN flake to act as a tunneling contact for the device.

Remote controlled micromanipulators in an Argon filled glovebox were used to align the transfer handle with the substrate chip such that roughly half of the monolayer MoS$_2$ overlaps the hBN or FLG/hBN. After contacting the hBN the transfer handle was slowly retracted until the MoS$_2$ tore at the edge of the hBN leaving half of it on the hBN and half of it on the transfer handle. Then the substrate chip was rotated by the desired twist angle using a rotation stage with a precision better than 0.01°. Then the other half of the MoS$_2$ was transferred on the first half creating a twisted MoS$_2$ on hBN (tMoS$_2$/hBN) stack.

In device C, an MLG was transferred on top the tMoS$_2$/hBN stack such that it partially covered the twisted region. This was done to reduce the contact resistance between the metallic electrode and the MoS$_2$ at low temperatures.

For each device, an electrode with a capacitance sensing geometry was defined using electron beam lithography and Titanium (4nm) and Gold (40 nm) were deposited in that order using an e-beam evaporator under UHV conditions.

Before loading the samples into the STM, they were annealed at 240 C in forming gas (10% Hydrogen + 90% Argon) for up to 16 hours to remove PMMA residues. The surface cleanliness was monitored between every sample fabrication step using Atomic Force Microscopy scans.

Table 1: Device summary

| Device | Stacking order | Target twist angle [$deg$] |
|---|---|---|
| A | P-tMoS2/MLG/hBN | 3 |
| B | P-tMoS2/MLG/hBN | 2 |
| C | MLG (partial)/P-tMoS2/hBN | 1 |
| RE-PFM | P-tMoS2/FLG/hBN | 0.2 |

STM/STS measurements:

STM measurements were performed using a home built STM system[27, 28, 42]. The tip was grounded while the bias was applied to the sample. Both chemically etched Tungsten tips as well as mechanically cut tips made from Platinum-Iridium wire were used. We noticed no difference in the measurements using either kind of tip. All tips were tested and processed on the gold electrode to verify sharpness and ensure a featureless density of states in the energy range of interest (+-3V). All topography scans were measured in constant-current mode. All dI/dV spectroscopy scans were collected in constant-height mode using a standard lock-in amplifier technique. An AC bias modulation of 15-20 mV rms was applied at frequencies of 4KHz or 797 Hz to measure spectroscopy data. Backgate voltage was applied using a battery bank and voltage divider circuit. All measurements were acquired at 77 K to avoid carrier freeze-out which occurs in these samples at lower temperatures

RE-PFM Measurements:

A twisted $MoS_2$/FLG/hBN stack with a designed twist angle of 0.2 was made using the same procedure used for making the STM devices. The stack was annealed overnight inside an argon-filled glovebox at 180 C to relax the structure. RE-PFM measurements were performed using an NT-MDT Solver Next AFM. Conducting probes with tip radius of curvature <10 nm and spring constant 12 N/m (K-Tek Nano HA_NC_Au) were used in contact mode. The contact resonance frequency was found to be in the 500-600 Hz range. An AC modulation up to 1V was applied to the probe at the contact resonance frequency while the sample was electrically floating. The vertical deflection signal of the laser spot and its phase with respect to the AC modulation were recorded.

Data Processing:

All topographic images were processed using Gwyddion[43]. Standard image processing techniques such as Median filtering and Gaussian smoothing were applied whenever necessary to reduce salt-and-pepper noise or bad data points associated with mechanical/acoustic noise during measurement.

All spectroscopic data represents the average of several curves acquired at the same position (typically 8 to 15 curves).

The mapping in Figure 4 was normalized to the dI/dV maximum for each curve which occurs near -2.4 V. Only the data in the range of the bias setpoint (-1.8 V onwards) is shown in Figure 4b for clarity.

The authors acknowledge support from the Department of Energy Grant No. DOE-FG02-99ER45742 and the Betty Moore Foundation EPiQS initiative Grant No. GBMF9453.

~

# Supplementary information for "Moiré Potential, Lattice Relaxation and Layer Polarization in Marginally Twisted MoS$_2$ Bilayers."

*Nikhil Tilak[1], Guohong Li[1], Takashi Taniguchi[2], Kenji Watanabe[2], Eva Y. Andrei[1]\**

1 Department of Physics and Astronomy, Rutgers, The State University of New Jersey, 136 Frelinghuysen Rd, Piscataway, NJ 08854, United States.

2 National Institute for Materials Science, Tsukuba, Japan.

Device Optical Microscope images and AFM scan.

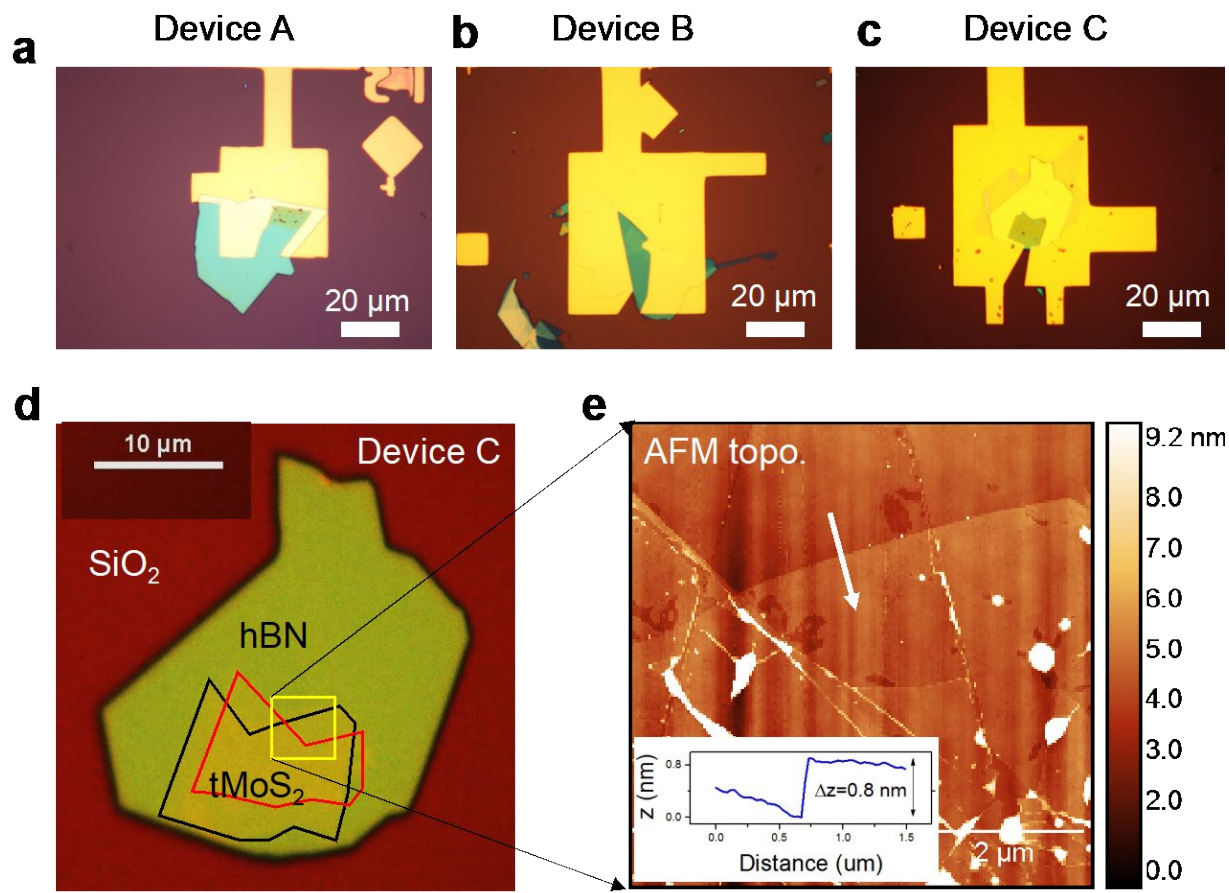

**Supplementary Figure 1**: (a-c) Optical microscope images of the three devices studied in this work. d The optical microscope image of the stack in Device C before the metal electrodes were fabricated. The bottom hBN and the two layers of monolayer MoS$_2$ can be seen. The overlapping area is the twisted MoS$_2$. e An Atomic Force topography scan gathered in tapping mode shows the region of Device C marked with a yellow square. The step height from the hBN to one of the MoS$_2$ layers is ~0.8 nm as seen in the section along the white arrow (inset).

Highly distorted topography due to strain

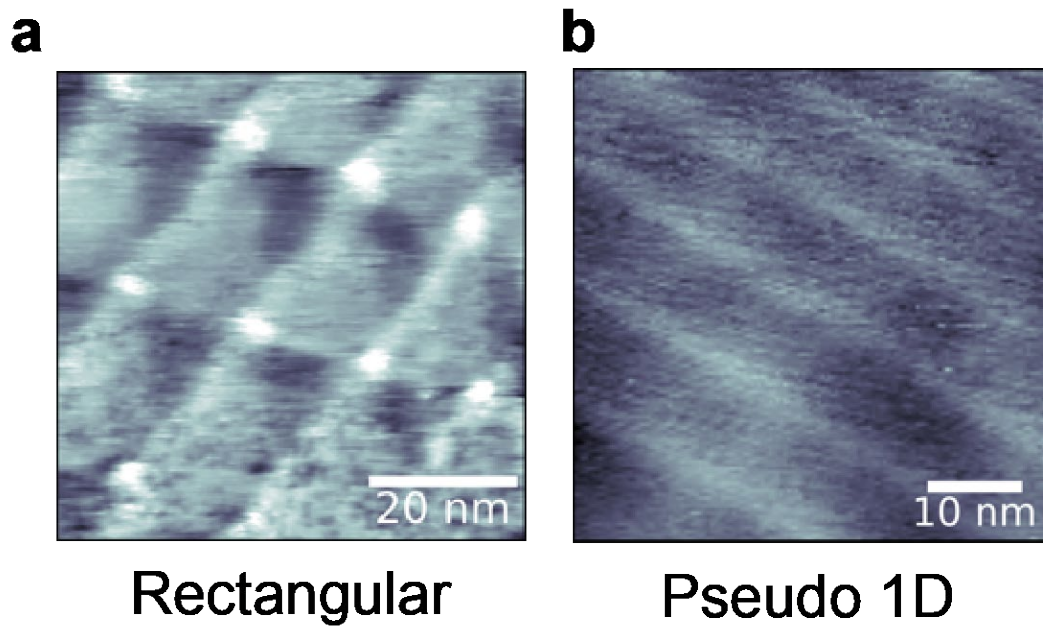

**Supplementary Figure 2**: Highly distorted moiré patterns due to strain. Examples of pseudo-rectangular (panel a) and pseudo-1D moiré patterns (panel b) observed in device C.

Height difference vs bias setpoint for the same current setpoint

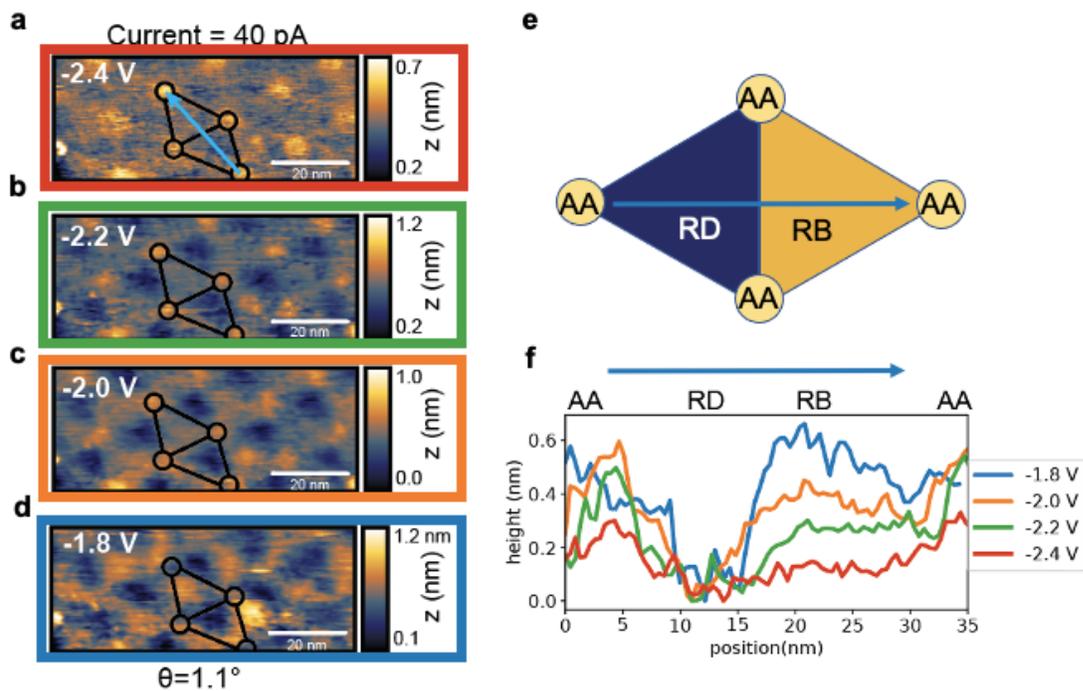

**Supplementary Figure 3**: (a-d) STM topography images of the same region taken at a current setpoint of 40 pA and bias setpoints of -2.4V, -2.2V, -2.0V and -1.8 V respectively. e A schematic drawing of a unit cell as seen in the topography scans. f the topographic height along the direction of the blue arrow in e. As the bias set point decreases, the topographic height difference between the bright (RB) and dark (RD) regions increases monotonically.

# Effects of Displacement field on MX/XM domains

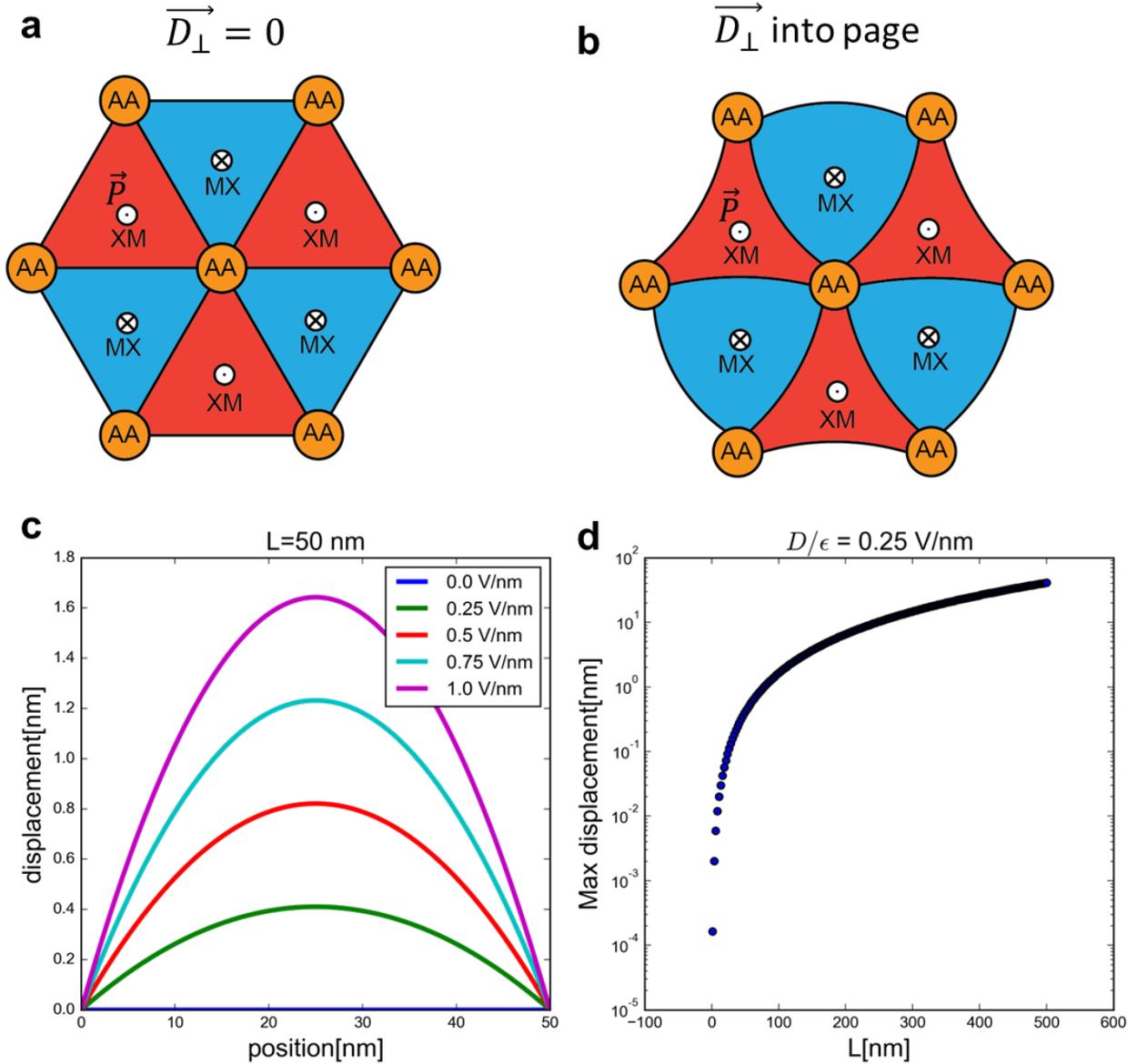

**Supplementary Figure 4**: a Illustration of the MX and XM domains with opposite signs of out-of-plane polarization. b shows the expansion of the MX domains via the bending of domain walls upon the application of a vertical displacement field pointing into the page. C. The bending of a domain wall of length 50 nm under the application of varying displacement fields. Values indicate

$\frac{D}{\epsilon} = \frac{D}{2.5\,\epsilon_0}$. d shows the maximum displacement of the domain walls under the application of $\frac{D}{\epsilon} = 0.25\ V/nm$. c and d were made using the model described in Weston et al[1].

Having established a direct link between the broken mirror symmetry in small angle P twisted MoS$_2$ and the broken layer symmetry, we attempted to perturb the system by applying a vertical displacement field using the Si backgate. This was motivated by two recent works which studied comparable samples using Piezoresponse Force Microscopy (PFM)[2] and back-scattered electron channeling contrast imaging (BSECCI)[1] respectively. The broken centrosymmetry in the MX and XM regions results in a spontaneous out-of-plane electric dipole which points in opposite directions in the MX and XM regions. In both studies it was shown that these ferroelectric domains could be made to grow or shrink in area by switching the magnitude and direction of a vertical displacement field. This was mediated via the motion of the domain walls as illustrated in Supplementary Figure 5a-b. This motion was also seen to be hysteretic in nature, i.e., once a large enough displacement field was applied, the domains locked into place even after the field is tuned back to 0. However, this was only possible for large triangular domains which are > 200 nm in size or in highly disordered regions where a periodic moiré lattice doesn't exist.

There are many reasons why it is difficult to achieve domain wall motion using STM. Firstly, while STM has excellent spatial resolution, it suffers from a relatively small field of view compared to AFM and BSECCI making it difficult to see large domains. For instance, the largest triangular domain size we observed was about 64 nm. Secondly, the vertical displacement field is highly non-uniform in our device as the electric field lines from the backgate terminate at the sharp STM tip. Thirdly, in back-gated devices the electric field due to the gate voltage is strongly screened by the bottom layer. As a consequence, we were unable to modify the MX/XM regions even after applying an estimated Displacement fields of $\frac{D}{\epsilon_0}$ ~ 0.6 V nm-1. Using the model described in Weston et al[1], we simulate the bending of a single domain wall of length 50 nm under various applied displacement fields (Supplementary Figure 5c). Note that even for unrealistically large Displacement fields the maximum displacement of the domain walls is ~ 1 nm. Due to the possibility of gate dielectric breakdown, we are constrained to a maximum Displacement field of $\frac{D}{\epsilon_0}$ ~ 0.6 V nm-1 (corresponding to Vg~80 V). For this displacement field we plotted the maximum displacement as a function of moire wavelength (Supplementary Figure 5d) which shows that for moire wavelengths <100 nm the maximum displacement of the domain walls remains < 1nm. This level of displacement is difficult to distinguish from effects such as strain. As a result, we did not see any noticeable changes in the sample topography as a function of Gate voltage.

## Large Area Resonance Enhanced PFM scans

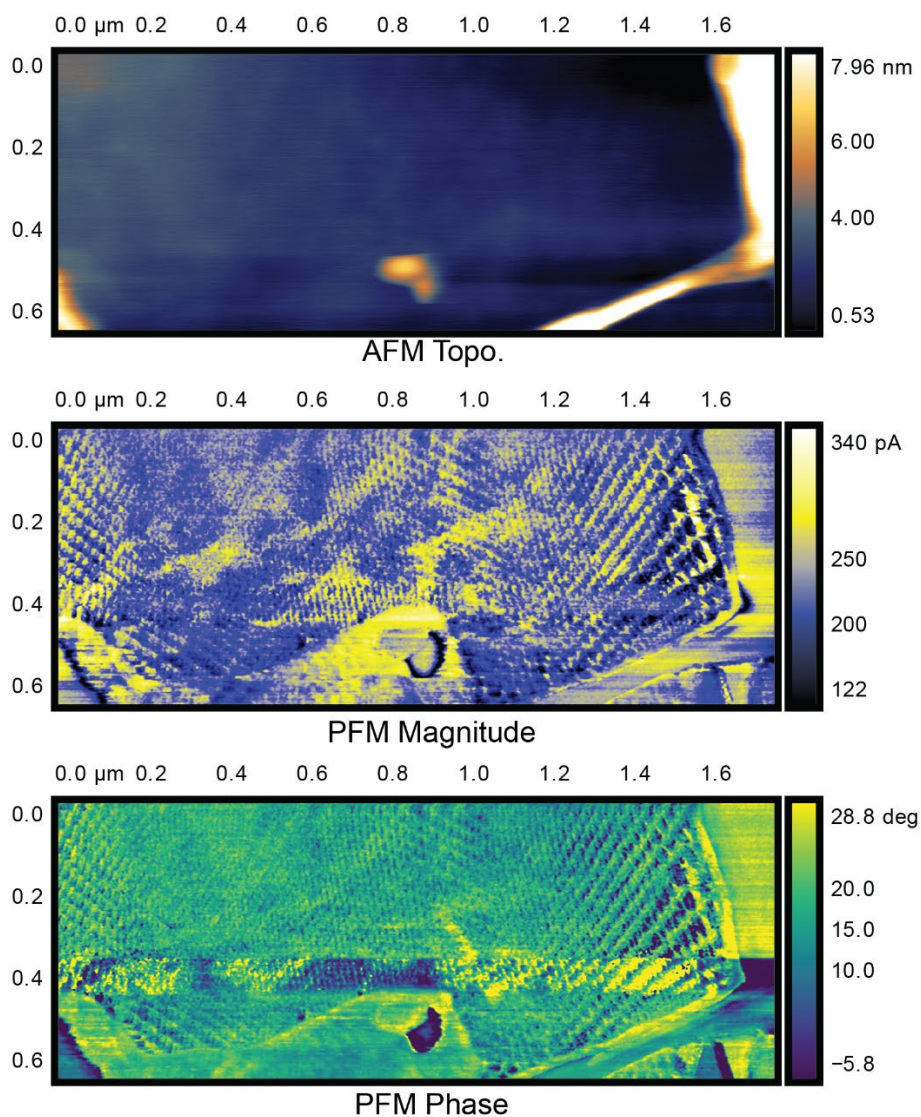

**Supplementary Figure 5:** Large area RE-PFM scans of a 0.2° twisted MoS$_2$ device. Topography, Magnitude and Phase scans from top to bottom. The tall regions in the topography scans are wrinkles created while assembling the tMoS2 stack.

In conventional PFM an AC electric field is applied between a conductive AFM tip and the sample and the resultant deformation of the sample surface is recorded via the deflection of the laser reflecting from the back of the cantilever. This technique produces magnitude and phase maps of the surface-deformation where domains with opposite vertical polarization have opposite contrast. Unfortunately, since the expected Piezoelectric coefficient for 3R-MoS$_2$ ($d_{33}$=0.3 pm/V[3]) is about 2 orders of magnitude smaller than conventional Piezoelectric materials such as Lead Zirconium Titanate (PZT)($d_{33}$~375 pm/V[4]) the resultant surface deformation would well below the resolving power of conventional PFM for moderate excitation amplitude. The deformation signal can be amplified by applying an AC modulation at the resonant frequency of the tip-sample contact. The effective vertical displacements get amplified by an amount equal to the Quality factor (Q~200) of this resonance. This is the so-called Resonance Enhanced PFM which we employed in this study (Main Figure 5 and Supplementary Figure 5).